\documentclass[12pt,aps,nofootinbib]{revtex4-2}

\usepackage{stmaryrd}
\usepackage{comment}
\usepackage{color}
\usepackage{amsmath}
\usepackage{amssymb}
\usepackage{amsthm}
\usepackage{mathrsfs}
\usepackage{graphicx}
\usepackage{marvosym}
\usepackage{amsmath,bm}
\usepackage{array}
\usepackage{latexsym}
\usepackage{enumerate}
\usepackage{slashed}
\usepackage{hyperref}
\usepackage{color}
\usepackage{setspace}
\usepackage[hang,small,bf]{caption}
\usepackage[subrefformat=parens]{subcaption}
\newcommand{\ri}{\mathrm{i}}
\newcommand{\rd}{\mathrm{d}}
\newcommand{\re}{\mathrm{e}}

\begin{document}
\title{Chiral kinetic theory with self-energy corrections \\ 
and neutrino spin Hall effect}

\author{Naoki Yamamoto}
\affiliation{Department of Physics, Keio University, Yokohama 223-8522, Japan}
\author{Di-Lun Yang}
\affiliation{Institute of Physics, Academia Sinica, Taipei, 11529, Taiwan}

\begin{abstract}
We systematically derive the chiral kinetic theory for chiral fermions with collisions, including the self-energy corrections, from quantum field theories. We find that the Wigner functions and chiral kinetic equations receive both the classical and quantum corrections from the self-energies and their spacetime gradients. We also apply this formalism to study nonequilibrium neutrino transport due to the interaction with thermalized electrons and nucleons, as realized in core-collapse supernovae. We derive neutrino currents along magnetic fields and neutrino spin Hall effect induced by the density gradient at first order in the Fermi constant $G_{\rm F}$ for anisotropic neutrino distributions.
\end{abstract}

\maketitle

\section{Introduction}

Recently, chiral kinetic theory (CKT)~\cite{Son:2012wh,Stephanov:2012ki,Son:2012zy,Chen:2012ca,Manuel:2013zaa,Manuel:2014dza,Chen:2015gta} has been established as a fundamental tool to describe nonequilibrium evolution of many-body ultrarelativistic fermions with chirality. The first-principles derivation of the CKT from the underlying quantum field theory via the Wigner function formalism~\cite{Hidaka:2016yjf,Hidaka:2017auj,Huang:2018wdl,Carignano:2019zsh} enables one to unambiguously determine the form of collisions in the CKT for a given microscopic theory. Based on this formalism, there have been extensive studies on the generalizations of the CKT, such as higher-order quantum corrections~\cite{Hayata:2020sqz,Mameda:2023ueq}, extensions to curved spacetime~\cite{Liu:2018xip,Yamamoto:2020zrs}, massive fermions~\cite{Gao:2019znl,Weickgenannt:2019dks,Hattori:2019ahi,Wang:2019moi,Yang:2020hri,Sheng:2021kfc,Manuel:2021oah}, and circularly polarized photons~\cite{Huang:2020kik,Hattori:2020gqh,Lin:2021mvw}; see Ref.~\cite{Hidaka:2022dmn} for a review. The CKT has been applied to various physical systems, such as quark-gluon plasmas in heavy ion collisions~\cite{Kharzeev:2015znc}, Weyl and Dirac semimetals~\cite{Armitage:2017cjs,Gorbar2021}, compact stars~\cite{Kamada:2022nyt}, and the early Universe~\cite{Kamada:2022nyt}. 

In the context of core-collapse supernovae where parity-violating effects can be relevant~\cite{Yamamoto:2015gzz}, the general relativistic form of the CKT for left-handed neutrinos with collisions has been systematically constructed based on the Standard Model of particle physics~\cite{Yamamoto:2020zrs} and applied to uncover novel chiral transport phenomena~\cite{Yamamoto:2021hjs,Matsumoto:2022lyb,Yamamoto:2022yva}. 
However, the previous CKT is not yet complete as it misses other potentially relevant chiral transport phenomena. One such example is the spin Hall effect of neutrinos induced by the density gradient, which has not been appreciated in the literature so far, to our knowledge. 
To derive a more generic Lorentz-covariant CKT that can describe the neutrino spin Hall effect and other chiral transport phenomena unexplored so far, further extension of the CKT is necessary by including the full self-energy corrections and quantum corrections systematically. Such an extension also modifies the free-streaming part and on-shell conditions of the CKT; in the case of the conventional kinetic theory, see Ref.~\cite{Blaizot:2001nr}.

In this paper, we derive the general CKT for chiral fermions with collisions incorporating such self-energy corrections from quantum field theories. We find that the self-energies and their spacetime gradients lead to both the classical and quantum corrections on the Wigner functions and chiral kinetic equations. Our main results are given by Eqs.~(\ref{eq:sol_WF})--(\ref{eq:C}). We then apply this formalism to study chiral transport phenomena of nonequilibrium neutrinos interacting with thermalized electrons. Such a situation is realized, e.g., in core-collapse supernovae, where the matter sector composed of electrons and nucleons is in thermal equilibrium due to the electromagnetic (or strong) interaction while neutrinos scattered with the matter sector only through the weak interaction are mostly out of equilibrium (see, e.g., Ref.~\cite{Bruenn:1985en}).
As a consequence, we find neutrino currents along magnetic fields and the neutrino spin Hall effect induced by the density gradient for anisotropic neutrino distributions; see Eqs.~(\ref{eq:eff_CME}) and (\ref{eq:J_SHE}), respectively.

We note that such a spin Hall effect is a universal feature of chiral particles not limited to neutrinos. In fact, it is known to appear also for photons~\cite{Liberman:1992zz,Bliokh:2004gz,Onoda:2004zz,Duval:2005ky,Gosselin:2006wp,Yamamoto:2017gla,Oancea:2020khc,Mameda:2022ojk} and gravitons~\cite{Yamamoto:2017gla,Andersson:2020gsj} with circular polarizations, e.g., in curved spacetime. 
A related spin Hall effect for quarks (as Dirac fermions) was also discussed in the context of heavy ion collisions~\cite{Liu:2020dxg}.

This paper is organized as follows. 
In Sec.~\ref{sec:Berry}, we provide a simple physical derivation of the neutrino spin Hall effect based on the notion of Berry curvature. In Sec.~\ref{sec:CKT}, we present the general formulation for the Wigner function and CKT for chiral fermions, including self-energies. In Sec.~\ref{sec:spin_Hall}, we apply the formalism to study neutrino transport due to the self-energy corrections and obtain the neutrino number current and energy-momentum tensor induced by magnetic fields and density gradient. Section~\ref{sec:summary} is devoted to summary and outlook. Technical details for the derivations are shown in appendices.

Throughout this paper, we focus on the flat spacetime. We use the mostly minus signature of the Minkowski metric $\eta^{\mu\nu} = {\rm diag} (1, -1,-1,-1)$ and the completely antisymmetric tensor $ \epsilon^{\mu\nu\alpha\beta}$ with $ \epsilon^{0123} = 1 $. The electric charge $e$ is absorbed into the definition of the gauge field $A^{\mu}$. We use the notations $A^{(\mu}B^{\nu)}\equiv A^{\mu}B^{\nu}+A^{\nu}B^{\mu}$ and $A^{[\mu}B^{\nu]}\equiv A^{\mu}B^{\nu}-A^{\nu}B^{\mu}$. We also define $\tilde{F}^{\mu\nu}\equiv\epsilon^{\mu\nu\alpha\beta}F_{\alpha\beta}/2$ with $F^{\mu\nu}$ being the electromagnetic field strength.

\section{Neutrino spin Hall effect from Berry curvature}
\label{sec:Berry}

In this section, we first provide a physical derivation of the neutrino spin Hall effect based on the semiclassical action for neutrinos including the effect of the Berry curvature, a notion widely applied in condensed matter physics~\cite{Xiao:2005qw}. 

We start with the generic semiclassical action for neutrinos \cite{Son:2012wh,Stephanov:2012ki,Son:2012zy,Yamamoto:2015gzz}:
\begin{align}
\label{action}
S = \int [{\bm p} \cdot \rd{\bm x} - (\epsilon_{\bm p} + V) \rd t - {\bm a}_{\rm p} \cdot \rd {\bm p}]\,,
\end{align}
where $\epsilon_{\bm p} = |\bm p|$ is the energy dispersion and $V$ is a generic potential energy (that will be specified in the context of supernovae later). As neutrinos are only left-handed  within the Standard Model, neutrinos have a nontrivial Berry curvature $\bm \Omega_{\bm p}$ in momentum space \cite{Yamamoto:2015gzz}:
\begin{align}
\bm \Omega_{\bm p} = - \hbar \frac{\bm p}{2 |\bm p|^3}\,.
\end{align}
This effect is incorporated in the action (\ref{action}) through the Berry connection ${\bm a}_{\bm p}$ in momentum space, which is related to the Berry curvature via $\bm \Omega_{\bm p} = {\bm \nabla} \times {\bm a}_{\bm p}$.

As a specific example of core-collapse supernovae, let us consider electron neutrinos in the electron and nucleon backgrounds. In this case, the backgrounds give rise to the potential energy \cite{Bethe:1986ej,Notzold:1987ik} (which we will also rederive in Sec.~\ref{sec:self_energy})
\begin{align}
\label{V}
V = \frac{G_{\rm F}}{\sqrt2} \left[(1+4\sin^2\theta_{\rm W}) N_{\rm e}
- N_{\rm n}
+ (1-4\sin^2\theta_{\rm W}) N_{\rm p} \right]
\,,
\end{align}
with $G_{\rm F}$ being the Fermi constant, $\theta_{\rm W}$ the Weinberg angle, and $N_{\rm e, n, p}$ the electron, neutron, and proton number densities.
The higher-order corrections of $\mathcal{O}(M_{W}^{-4})$ in $V$~\cite{Notzold:1987ik} are negligible in the regime $m_{W}^2 \gg E_{\nu} E_{\rm e}$ \cite{Shukla:1997yx,Shukla1999} in the context of supernovae, where $m_{W}$ is the mass of $W$ bosons, $E_{\nu}$ is the neutrino energy, and $E_{\rm e}$ is the electron energy.
It is well known that the potential energy $V$ leads to the so-called Mikheyev-Smirnov-Wolfenstein effect in the context of neutrino oscillations \cite{Wolfenstein:1977ue,Mikheyev:1985zog}, but here we point out yet another medium effect in neutrino physics---neutrino spin Hall effect.

The semiclassical equation of motion for neutrinos follow from the action (\ref{action}) as
\begin{align}
\label{xdot}
\dot {\bm x} &= {\bm v} + \dot {\bm p} \times {\bm \Omega}_{\bm p},
\\
\label{pdot}
\dot {\bm p} &= - {\bm \nabla} V,
\end{align}
where ${\bm v} = \partial \epsilon_{\bm p}/ \partial {\bm p}$ is the velocity of neutrinos. When the densities $N_{\rm e,n,p}$ vary depending on $x$, the variation of the potential energy $V$ leads to the force given by the right-hand side of Eq.~(\ref{pdot})~\cite{Shukla1999}. The new ingredient compared with the previous literature in our formulation is the contribution in Eq.~(\ref{xdot}) expressed by the Berry curvature. This gives rise to the additional contribution in the neutrino number current:
\begin{align}
\label{eq:SHE}
{\bm J} = \int \frac{\rd^3 {\bm p}}{(2\pi)^3} \dot {\bm x} f^{(\nu)}
\supset - \int \frac{\rd^3 {\bm p}}{(2\pi)^3} {\bm \nabla} V \times {\bm \Omega}_{\bm p} f^{(\nu)} =: {\bm J}_{\rm SHE}\,,
\end{align}
where $f^{(\nu)}(t, {\bm p}, {\bm x})$ is the neutrino distribution function in the phase space. This current flowing in the direction perpendicular to ${\bm \nabla} V$ is the neutrino spin Hall effect, by analogy with the conventional spin Hall current triggered by a transverse electric field. This should be contrasted with the classical current in the direction parallel to ${\bm \nabla} V$. To our knowledge, this is the first to demonstrate the neutrino spin Hall effect. Note that this current can be nonvanishing only when the momentum distribution of neutrinos is anisotropic. 

This derivation of the neutrino spin Hall effect is based on the semiclassical description of neutrinos in the phase space, augmented by the effect of the Berry curvature.
However, a drawback of the resulting kinetic theory is that it lacks a manifest Lorentz covariance and cannot be extended, e.g., to curved spacetime. Using the Wigner function formalism and systematically including the self-energy and quantum corrections, such a Lorentz-covariant kinetic theory can be derived from the underlying quantum field theory.
By doing so, we will obtain the generic CKT that not only reproduces the neutrino spin Hall effect, but delineates other chiral transport phenomena, as shown below.

\section{Chiral kinetic theory with self-energy corrections}
\label{sec:CKT}
In this section, we shall systematically derive the generic CKT with self-energy corrections associated with the modifications of the dispersion relation and those contributing to collision terms based on Wigner functions and $\hbar$ expansion.
In the conventional power counting scheme, the gradient correction in phase space is considered as the same order as self-energies depending on the interaction at weak coupling. This can be understood from simply a classical Boltzmann equation, for which the gradient term in the free-streaming part can be balanced by the collision term without gradient corrections. In general, the gradient expansion and the coupling-constant expansion can be separated. In the following, we keep $\hbar$ only to characterize the gradient correction and set $\hbar = 1$ for other corrections in the coupling-constant expansion. 

The central object in the formulation is the lesser and greater Wigner functions for chiral fermions $\psi_{\chi}$ with chirality $\chi = {\rm R,L}$:
\begin{align}
\mathcal{W}_{\chi}^{<}(q,x)&\equiv\int {\rm d}^4y {\rm e}^{-{\rm i}q\cdot y}\langle \psi^{\dagger}_{\chi}(x+y/2)\psi_{\chi}(x-y/2)\rangle \,, \\
\mathcal{W}_{\chi}^{>}(q,x)&\equiv\int {\rm d}^4y {\rm e}^{-{\rm i}q\cdot y}\langle\psi_{\chi}(x-y/2){\psi}^{\dagger}_{\chi}(x+y/2)\rangle \,.
\end{align}
Here the gauge link is implicitly embedded, and $q^{\mu}$ represents the kinetic momentum.
The Kadanoff-Baym equations for $\mathcal{W}_{\chi}^{<}$ up to $\mathcal{O}(\hbar)$ read~\cite{Hidaka:2022dmn}
\begin{align}
\label{eq:KB_R}
\sigma^{\mu}\left(q_{\mu}+\frac{1}{2}\ri\hbar\Delta_{\mu}\right)\mathcal{W}^{<}_{\rm R}-\bar{\Sigma}_{\rm R}\star \mathcal{W}^{<}_{\rm R}&=\frac{\ri}{2}\left(\Sigma^<_{\rm R}\star\mathcal{W}^{>}_{\rm R}-\Sigma^>_{\rm R}\star\mathcal{W}^{<}_{\rm R}\right)\,, \\
\label{eq:KB_L}
\bar{\sigma}^{\mu}\left(q_{\mu}+\frac{1}{2}\ri\hbar\Delta_{\mu}\right)\mathcal{W}^{<}_{\rm L}-\bar{\Sigma}_{\rm L}\star \mathcal{W}^{<}_{\rm L}&=\frac{\ri}{2}\left(\Sigma^<_{\rm L}\star\mathcal{W}^{>}_{\rm L}-\Sigma^>_{\rm L}\star\mathcal{W}^{<}_{\rm L}\right)\,,
\end{align}
where $\sigma^{\mu}=(1,{\bm \sigma})$ and $\bar \sigma^{\mu}=(1, -{\bm \sigma})$ with $\sigma^i$ ($i=1,2,3$) being Pauli matrices, $\Delta_{\mu}=\partial_{\mu}+F_{\nu\mu}\partial^{\nu}_{q}$, 
$\bar{\Sigma}_{\chi}={\rm Re}(\Sigma^{{\rm r}}_{\chi })+\Sigma^{\delta}_{\chi}$ with $\Sigma^{{\rm r}}_{\chi}$ being the retarded self-energy%
\footnote{In the real-time formalism, we adopt the convention~\cite{Hidaka:2022dmn} $\Sigma^{\rm r} = \Sigma^{++}-\Sigma^{+-}=\Sigma^{-+}-\Sigma^{--}$, $\Sigma^{\rm a} = \Sigma^{++}-\Sigma^{-+}=\Sigma^{+-}-\Sigma^{--}$, $\ri\Sigma^{+-} = -\Sigma^{<}$, and 
$\ri\Sigma^{-+} = \Sigma^{>}$, where the superscript $+,-$ denotes the time branch on the closed time path and the subscript $\chi={\rm R},{\rm L}$ for chirality is omitted for brevity.} 
and $\Sigma^{\delta}_{\chi}$ the one-point potential, $\Sigma^{\lessgtr}_{\chi}$ are the lesser and greater self-energies, and the operator $\star$ is the Moyal product.
In general, the self-energy $\Sigma$ is decomposed into right- and left-handed  components as $\Sigma=P_{\rm L}\gamma^{\mu}\Sigma_{{\rm R}\mu}+P_{\rm R}\gamma^{\mu}\Sigma_{{\rm L}\mu}$ with $P_{\rm R}=(1+\gamma^5)/2$ and $P_{\rm L}=(1-\gamma^5)/2$. 

The Wigner functions and self-energies can be decomposed as $\mathcal{W}^{<}_{\rm R}=\bar{\sigma}_{\mu}\mathcal{W}^{<\mu}_{\rm R}$, $\mathcal{W}^{<}_{\rm L}=\sigma_{\mu}\mathcal{W}^{<\mu}_{\rm L}$, $\bar{\Sigma}_{\rm R}=\sigma_{\mu}\bar{\Sigma}^{\mu}_{\rm R}$, $\bar{\Sigma}_{\rm L}=\bar{\sigma}_{\mu}\bar{\Sigma}^{\mu}_{\rm L}$, $\Sigma^{\lessgtr}_{\rm R}=\sigma_{\mu}\Sigma^{\lessgtr\mu}_{\rm R}$, and $\Sigma^{\lessgtr}_{\rm L}=\bar{\sigma}_{\mu}\Sigma^{\lessgtr\mu}_{\rm L}$. To write down the Kadanoff-Baym equations for $\mathcal{W}^{<\mu}_{\chi}$ with chirality $\chi={\rm R},{\rm L}$, we use the relations
\begin{eqnarray}
\sigma^{\mu}\bar{\sigma}^{\nu}=\eta^{\mu\nu}-n^{[\mu}\sigma^{\nu]}+\ri\epsilon^{\mu\nu\alpha\beta}n_{\alpha}\sigma_{\beta},\quad
\bar{\sigma}^{\mu}\sigma^{\nu}=\eta^{\mu\nu}+n^{[\mu}\sigma^{\nu]}+\ri\epsilon^{\mu\nu\alpha\beta}n_{\alpha}\sigma_{\beta},
\end{eqnarray}
where the timelike frame vector $n^{\mu}$ satisfying $n^2=1$ is introduced to specify the choice of the spin basis. Equations~(\ref{eq:KB_R}) and (\ref{eq:KB_L}) then lead to
\begin{gather}
\left(q_{\mu}+\frac{\ri\hbar}{2}\Delta_{\mu} -\bar{\Sigma}_{\chi\mu}\star \right)\mathcal{W}^{<\mu}_{\chi}
=\frac{\ri}{2}\left(\Sigma^{<\mu}_{\chi}\star\mathcal{W}^{>}_{\chi\mu}-\Sigma^{>\mu}_{\chi}\star\mathcal{W}^{<}_{\chi\mu}\right)\,,
\\
\sigma_{\perp\beta}(-\chi n^{[\mu}\eta^{\nu]\beta}+\ri\epsilon^{\mu\nu\alpha\beta}n_{\alpha})\bigg[\left(q_{\mu}+\frac{\ri\hbar}{2}\Delta_{\mu} -\bar{\Sigma}_{\chi\mu}\star \right)\mathcal{W}^{<}_{\chi\nu}
-\frac{\ri}{2}\left(\Sigma^{<}_{\chi\mu}\star\mathcal{W}^{>}_{\chi\nu}-\Sigma^{>}_{\chi\mu}\star\mathcal{W}^{<}_{\chi\nu}\right)
\bigg]=0\,,
\end{gather}
where the coefficient $\chi=\pm 1$ corresponds to the subscript $\chi={\rm R,L}$ for chirality. Here, we introduced the notation $V^{\mu}_{\perp}\equiv (\eta^{\mu\nu}-n^{\mu}n^{\nu})V_{\nu}$ for an arbitrary four-vector $V^{\mu}$. 

Using the $\hbar$ expansion of the Moyal product for generic $A(q,x)$ and $B(q,x)$, 
\begin{eqnarray}
A{\star}B=AB+\frac{\ri\hbar}{2}\{A,B\}_{\rm PB}-\frac{\ri\hbar}{2}F_{\mu\nu}\partial^{\mu}_q A \partial^{\nu}_q B + \mathcal{O}(\hbar^2)\,,
\end{eqnarray}
where $\{A,B\}_{\rm PB}=(\partial^{\nu}_{q}A)(\partial_{\nu}B)-(\partial_{\nu}A)(\partial^{\nu}_{q}B)$ denotes the Poisson bracket, the master equations obtained from the Kadanoff-Baym equations up to $\mathcal{O}(\hbar)$ are given by
\begin{gather}
\label{eq:master_KE}
\tilde{\mathcal{D}}_{\mu}\mathcal{W}^{<\mu}_{\chi}=0, 
\\
\label{eq:master_onshellcond}
\tilde{q}_{\mu}\mathcal{W}^{<\mu}_{\chi}=0,
\\
\label{eq:master_auxiliary}
\quad \tilde{q}^{[\nu}\mathcal{W}^{<\mu]}_{\chi}
=\chi\frac{\hbar}{2}\epsilon^{\mu\nu\rho\sigma}\tilde{\mathcal{D}}_{\rho}\mathcal{W}^{<}_{\chi\sigma},
\end{gather}
where we introduce 
$\mathcal{D}_{\rho}\mathcal{W}^{<}_{\chi\sigma}=\Delta_{\rho}\mathcal{W}^{<}_{\chi\sigma}-\Sigma^{<}_{\chi\rho}\mathcal{W}^{>}_{\chi\sigma}+\Sigma^{>}_{\chi\rho}\mathcal{W}^{<}_{\chi\sigma}$, $\tilde{\mathcal{D}}_{\rho}=\mathcal{D}_{\rho}+(\Delta_{\nu}\bar{\Sigma}_{\chi\rho})\partial^{\nu}_{q}-(\partial_{q\nu}\bar{\Sigma}_{\chi\rho})\partial^{\nu}$, and $\tilde{q}_{\mu}=q_{\mu}-\bar{\Sigma}_{\chi\mu}$.
Equations~(\ref{eq:master_onshellcond}) and (\ref{eq:master_auxiliary}) determine the perturbative solution of the Wigner function, while Eq.~(\ref{eq:master_KE}) leads to the kinetic equation.

To solve Eqs.~(\ref{eq:master_onshellcond}) and (\ref{eq:master_auxiliary}) for the Wigner functions perturbatively, 
we make the decomposition $\mathcal{W}^{<\mu}_{\chi}=\mathcal{W}^{(0)<\mu}_{\chi}+\hbar \mathcal{W}^{(1)<\mu}_{\chi}$. For simplicity, we will drop nonlinear terms ${\cal O}(\bar{\Sigma}_{\chi}^2)$, ${\cal O}(\bar{\Sigma}_{\chi}\Sigma^{\lessgtr}_{\chi})$, and ${\cal O}\big((\Sigma^{\lessgtr}_{\chi})^2 \big)$ at weak coupling. Here, we present crucial steps for the derivation, while more technical details are shown in Appendix~\ref{app:KBeq_solution}. From Eqs.~(\ref{eq:master_onshellcond}) and (\ref{eq:master_auxiliary}), the leading-order solution takes the form
\begin{eqnarray}
\label{eq:solution_leading}
	\mathcal{W}^{(0)<\mu}_{\chi}(q,x)=2\pi {\rm sgn}(q_0)\delta(\tilde{q}^2)\tilde{q}^{\mu}f_{\chi},
\end{eqnarray} 
where $f_{\chi}$ denotes the distribution function for chiral fermions and ${\rm sgn}(q_0)$ represents the sign of $q_0$, which is introduced to incorporate the contributions from both particles and antiparticles. Here $\delta(\tilde{q}^2)$ characterizes the on-shell condition with the self-energy corrections. By plugging $\mathcal{W}^{(0)<\mu}_{\chi}$ into Eq.~(\ref{eq:master_KE}), the corresponding kinetic equation up to $\mathcal{O}(\hbar^0)$ reads
\begin{eqnarray}
\label{KE_leading}
\delta(\tilde{q}^2)\tilde{q}^{\mu}\big(\tilde{\Delta}_{\mu}f_{\chi}-\mathcal{C}_{\mu}[f_{\chi}]\big)
=0,
\end{eqnarray}
where $\tilde{\Delta}_{\rho}=\Delta_{\rho}+(\Delta_{\nu}\bar{\Sigma}_{\chi\rho})\partial^{\nu}_{q}-(\partial_{q\nu}\bar{\Sigma}_{\chi\rho})\partial^{\nu}$ and $\mathcal{C}_{\mu}[f_{\chi}]=\Sigma^{<}_{\chi\mu}(1-f_{\chi})-\Sigma^{>}_{\chi\mu}f_{\chi}$ represents the collision term.

Subsequently, the next-to-leading-order correction $\mathcal{W}^{(1)<\mu}_{\chi}$ has to be obtained by solving 
\begin{eqnarray}
\label{eq:W1}
\tilde{q}^{[\nu}\mathcal{W}^{(1)<\mu]}_{\chi}
	=\frac{\chi}{2}\epsilon^{\mu\nu\rho\sigma}\tilde{\mathcal{D}}_{\rho}\mathcal{W}^{(0)<}_{\chi\sigma}
\end{eqnarray}
with the constraint $\tilde{q}_{\mu}\mathcal{W}^{(1)<\mu}_{\chi}=0$. Making contraction of Eq.~(\ref{eq:W1}) with $\tilde{q}_{\nu}$ and $n_{\nu}$, one finds
\begin{gather}
\label{eq:W1_q}
	\tilde{q}^{2}\mathcal{W}^{(1)<\mu}_{\chi}
	=\frac{\chi}{2}\epsilon^{\mu\nu\rho\sigma}\tilde{q}_{\nu}\tilde{\Delta}_{\rho}\mathcal{W}^{(0)<}_{\chi\sigma}\,,
	\\
\label{eq:W1_n}
	\tilde{q}\cdot n\mathcal{W}^{(1)<\mu}_{\chi}-n\cdot\mathcal{W}^{(1)<}_{\chi}\tilde{q}^{\mu}
	=\frac{\chi}{2}\epsilon^{\mu\nu\rho\sigma}n_{\nu}\tilde {\cal D}_{\rho} {\cal W}^{(0)<}_{\chi\sigma}\,,
\end{gather}
respectively.
From these two equations, one can derive the solution of $\mathcal{W}^{(1)<\mu}_{\chi}$:
\begin{eqnarray}
\label{eq:W1_solution}
	\mathcal{W}^{(1)<\mu}_{\chi}
	=2\pi\chi {\rm sgn}(q_0)\left[\delta(\tilde{q}^2)S^{\mu\nu}_{\tilde{q}}\big(\tilde{\Delta}_{\nu}f_{\chi}-\mathcal{C}_{\nu}[f_{\chi}]\big)
	+\frac{1}{2}\delta'(\tilde{q}^2)\epsilon^{\mu\nu\rho\sigma}\tilde{q}_{\nu}(\tilde{\Delta}_{\sigma}\tilde{q}_{\rho})f_{\chi}\right]\,,
\end{eqnarray} 
where $S^{\mu\nu}_{\tilde{q}}\equiv \epsilon^{\mu\nu\alpha\beta}\tilde{q}_{\alpha}n_{\beta}/(2\tilde{q}\cdot n)$ is the spin tensor modified by the presence of $\bar{\Sigma}_{\chi\mu}$ and $\delta'(x)\equiv\partial_{x}\delta(x)$. 
The perturbative solution up to $\mathcal{O}(\hbar)$, $\mathcal{O}(\bar{\Sigma}_{\chi})$, and $\mathcal{O}(\Sigma^{\lessgtr}_{\chi})$ is accordingly given by 
\begin{eqnarray}
\label{eq:sol_WF}
	\mathcal{W}^{<\mu}_{\chi}=2\pi {\rm sgn}(q_0)\left[\delta(\tilde{q}^2)\big(\tilde{q}^{\mu}+\chi\hbar S^{\mu\nu}_{\tilde{q}}\tilde{\mathcal{D}}_{\nu}\big)
	+ \frac{\chi\hbar}{2} \delta'(\tilde{q}^2) \epsilon^{\mu\nu\rho\sigma}\tilde{q}_{\nu}\big(F_{\rho\sigma}+\Delta_{[\rho}\bar{\Sigma}_{\chi\sigma]}\big)\right]f_{\chi}\,.
	\nonumber \\
\end{eqnarray} 
In fact, one can check that Eq.~(\ref{eq:sol_WF}) satisfies Eq.~(\ref{eq:W1}) as shown in Appendix~\ref{app:KBeq_solution}.  

Given the perturbative solution, with complicated yet straightforward computations, Eq.~(\ref{eq:master_KE}) results in the corresponding chiral kinetic equation up to $\mathcal{O}(\hbar)$,  $\mathcal{O}(\bar{\Sigma}_{\chi})$, and $\mathcal{O}(\Sigma^{\lessgtr}_{\chi})$:
\begin{eqnarray}
\label{eq:offshell_CKE}\nonumber
&&\bigg(\tilde{q}^{\nu}+\frac{\chi\hbar S^{\mu\nu}_{\tilde{q}}}{\tilde{q}\cdot n}\left[n^{\alpha}(\Delta_{[\mu}\bar{\Sigma}_{\chi\alpha]})+E_{\mu}-\tilde{q}^{\alpha}(\tilde{\Delta}_{\mu}n_{\alpha}) \right]
+\frac{\chi\hbar\epsilon^{\mu\nu\rho\sigma}\tilde{q}_{\rho}(\tilde{\Delta}_{\mu}n_{\sigma})}{2\tilde{q}\cdot n}
+\chi\hbar S^{\mu\nu}_{\tilde{q}}\tilde{\Delta}_{\mu}\bigg) \tilde{\Delta}_{\nu}
f_{\chi} \\
&&=C[f_{\chi}] 
\end{eqnarray}
with the on-shell condition $\tilde{q}^2=-\chi\hbar S^{\mu\nu}_{\tilde{q}}(\tilde{\Delta}_{\mu}\tilde{q}_{\nu})$ for the spacetime-dependent frame vector $n^{\mu}(x)$, where
$\tilde{\Delta}_{\mu}n_{\alpha}=\partial_{\mu}n_{\alpha}-(\partial_q^{\nu}\bar{\Sigma}_{\chi \mu})\partial_{\nu}n_{\alpha}$ and 
\begin{eqnarray}
\label{eq:C}
	C[f_{\chi}]=\left[q^{\nu}+\frac{\chi\hbar S^{\mu\nu}_qE_{\mu}}{q\cdot n}+\chi\hbar(\partial_{\mu} S^{\mu\nu}_q)\right]\mathcal{C}_{\nu}[f_{\chi}]+\frac{\chi\hbar\epsilon^{\mu\nu\alpha\beta}}{2q\cdot n} q_{\mu}n_{\nu} \big[f_{\chi}\Delta_{\alpha}\Sigma^{>}_{\chi\beta}-(1-f_{\chi})\Delta_{\alpha}\Sigma^{<}_{\chi\beta}\big]
	\nonumber \\
\end{eqnarray}
denotes the collision term with $\hbar$ corrections shown, e.g., in Ref.~\cite{Kamada:2022nyt}. Here the electric and magnetic fields are defined in the frame specified by $n^{\mu}$ as
$F_{\mu\nu}n^{\nu}=E^{\mu}$ and $\tilde{F}_{\mu\nu}n^{\nu}=B^{\mu}$.
This is one of the main results of this paper. The formalism presented here completes the previous one in Ref.~\cite{Hidaka:2016yjf} without the contribution from $\bar{\Sigma}_{\chi}$.

\section{Magnetic-field corrections and spin Hall effect in neutrino transport}\label{sec:spin_Hall}
In this section, we will apply the CKT derived in the previous section to investigate the self-energy ($\bar{\Sigma}_{\rm L}$) corrections on the Wigner functions and chiral kinetic equation of left-handed neutrinos. We shall focus on the medium effect due to thermalized electrons under a homogeneous magnetic field. As will be shown later, the magnetic field gives rise to further modifications on the kinetic equation of neutrinos on top of the $\hbar$ corrections in the collision terms previously found in Ref.~\cite{Yamamoto:2020zrs}. Moreover, for electrons in local equilibrium, we derive the neutrino spin Hall current perpendicular to the density gradient for anisotropic neutrino distributions in momentum space.

\subsection{Calculation of the retarded self-energy}
\label{sec:self_energy}
\begin{figure}[h]
\centering
\includegraphics[width=11cm]{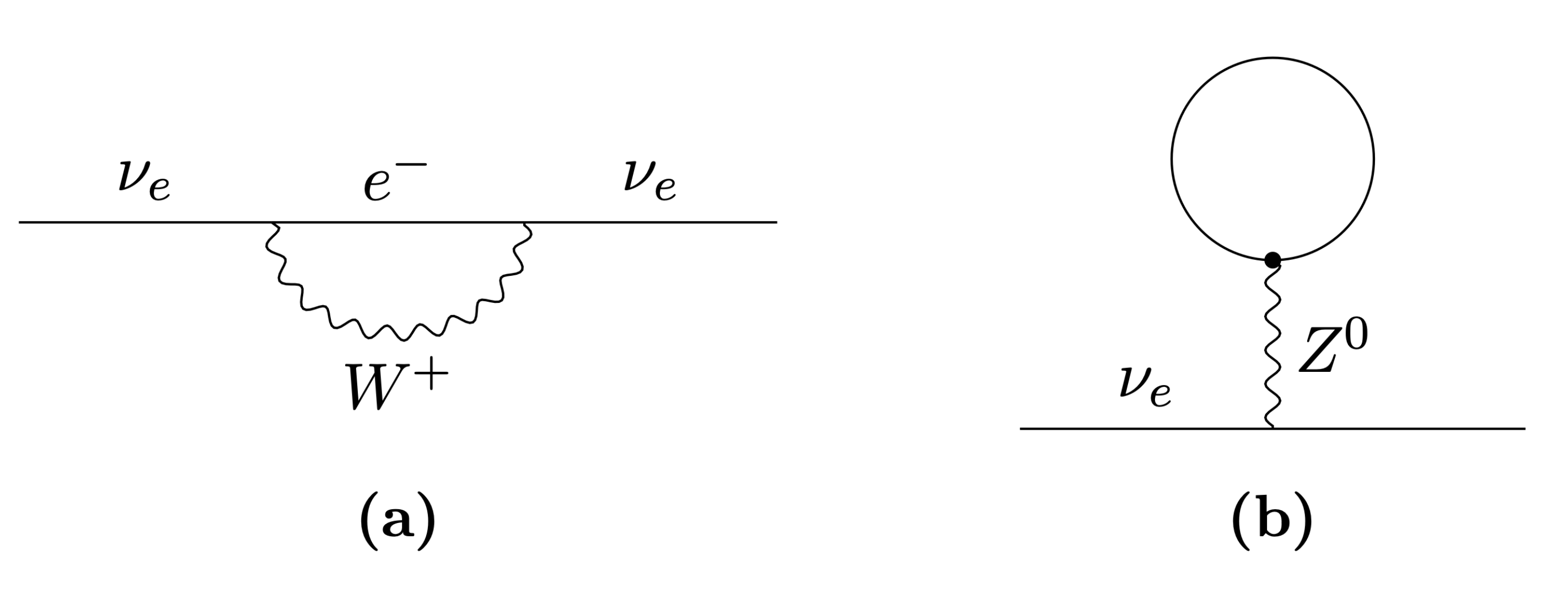}
  \caption{One-loop contributions to the neutrino self-energy: (a) bubble diagram via the exchange of the $W^+$ boson and (b) tadpole diagram via the exchange of the $Z^0$ boson.}
  \label{fig:feynman}
\end{figure}

We shall begin with the calculation of $\bar{\Sigma}_{\rm L}$ for left-handed electron neutrinos. The neutrino self-energies in medium without and with magnetic fields were previously studied in Refs.~\cite{Notzold:1987ik} and ~\cite{DOlivo:1989ued,Esposito:1995db,Elizalde:2000vz,Elizalde:2004mw}, respectively. For example, the lowest Landau level approximation for a strong and constant magnetic field is adopted in Ref.~\cite{Elizalde:2004mw}.
Here, we adopt a different treatment and approximation from previous works and consider a relatively weak magnetic field such that it is treated as a derivative expansion in the Wigner function formalism.
From the weak interaction, ${\rm Re}(\Sigma^{{\rm r}\mu}_{\rm L })$ and $\Sigma^{\delta \mu}_{\rm L}$ are derived from the one-loop contributions shown in Figs.~\ref{fig:feynman}(a) and \ref{fig:feynman}(b), respectively. In the following, let us first focus on the contribution from Fig.~\ref{fig:feynman}(a) and then the one from Fig.~\ref{fig:feynman}(b) in the electron background, and finally the one from Fig.~\ref{fig:feynman}(b) in the nucleon background.

Considering the charged-current interaction 
\begin{eqnarray}
\mathcal{L}_{\rm cc}=\frac{g}{\sqrt{2}}\bar{\psi}_{\rm e}\gamma_{\mu}P_{\rm L}\mathcal{A}^{\mu}_{ W^+}\psi_{\nu}+{\rm H.c.},
\end{eqnarray}
where $\mathcal{A}^{\mu}_{W^+}$ denotes the gauge field for $W^+$ bosons, one finds the contribution in the electron background%
\footnote{Here the prefactor ``$\ri$" in front of $\Sigma^{++}_{\rm L,e}(q)$ is introduced for convention. As will be shown in the following computations, the sign in front of ``$\ri$" is chosen to obtain a positive correction on $q^2$ for the on-shell condition.} 
\begin{eqnarray}\label{eq:iSigmaL_cal}
\ri\Sigma^{++}_{\rm L,e}(q)=\frac{g^2}{2}\int \frac{\rd^4p}{(2\pi)^4}\big[\gamma^{\alpha} P_{\rm L} S^{({\rm e})}(p)\gamma^{\beta} P_{\rm L} \big]G^{W}_{\alpha\beta}(q-p)\,,
\end{eqnarray}
where $S^{({\rm e})}(q)$ and $G^{W}_{\alpha\beta}(q)$ are the Feynman propagators of electrons and $W^+$ bosons, respectively. Accordingly, we have
\begin{eqnarray}
\ri\Sigma^{++\mu}_{\rm L,e}(q)=\frac{g^2}{4}\int \frac{\rd^4p}{(2\pi)^4}{\rm Tr}\big[\gamma^{\mu}\gamma^{\alpha}S^{({\rm e})}_{\rm L}(p)\gamma^{\beta}\big]G^{W}_{\alpha\beta}(q-p)\,,
\end{eqnarray}
where $S^{({\rm e})}_{\chi}(q)=P_{\chi}\gamma^{\mu}\mathcal{W}_{{\chi}\mu}(q)$. 
Considering the low-energy regime where $q^2 \ll M_{W}^2$, we can approximate
\begin{eqnarray}
G^{W}_{\alpha\beta}(q)=\frac{-\ri\eta_{\alpha\beta}}{q^2-M_{W}^2}\approx \frac{\ri\eta_{\alpha\beta}}{M_{W}^2}\,,
\end{eqnarray}
so that
\begin{eqnarray}
{\rm Tr}\big[\gamma^{\mu}\gamma^{\alpha}S^{({\rm e})}_{\rm L}(p)\gamma^{\beta}\big]G^{W}_{\alpha\beta}(q-p)
=-\frac{4\ri}{M_W^2}\mathcal{W}^{\mu}_{\rm L}(p)\,.
\end{eqnarray}

As presented in Appendix~\ref{app:FeynmannW}, the explicit form of $\mathcal{W}^{\mu}_{\rm L}$ for left-handed electrons in thermal equilibrium with a constant magnetic field is given by%
\footnote{For electrons in local equilibrium, $\mathcal{W}^{\mu}_{\rm L}$ can, in principle, include the $\hbar$ corrections of the temperature and chemical potential gradients and vorticity. We omit these corrections for simplicity.}
\begin{eqnarray}
\label{W_B}
	\mathcal{W}^{\mu}_{\rm L}(q)=\frac{\ri q^{\mu}}{q^2+\ri\epsilon}
	+\hbar \left[\frac{\ri q_0B^{\mu}}{(q^2)^2}+\pi B^{[\mu}n^{\nu]}q_{\nu}\delta'(q^2)\right]
	-2\pi \left[q^{\mu}\delta(q^2)-\hbar B^{[\mu}n^{\nu]}q_{\nu}\delta'(q^2)\right]\tilde f_{\rm L}(q)\,,
	\nonumber \\
\end{eqnarray}
where 
\begin{align}
\label{n_L}
\tilde f_{\rm L}(q)=\frac{\Theta(q_0)}{\re^{\beta(|q_0|-\mu_{\rm e})}+1}+\frac{\Theta(-q_0)}{\re^{\beta(|q_0|+\mu_{\rm e})}+1}\,,
\end{align}
with $\beta=1/T$ and $\mu_{\rm e}$ being the local inverse temperature and chemical potential of left-handed electrons, respectively. 
For right-handed electrons, the $\hbar$ terms should flip the signs. Hereafter we work with $n^{\mu}=\bar{n}^{\mu}\equiv (1,\bm 0)$. 
Also, as we always consider left-handed fermions, we will omit the subscript ``L" for left-handedness unless specified. 

Since we are interested in the medium contribution from electrons in thermal equilibrium, we may simply input the $T$- and/or $\mu$-dependent parts of $\mathcal{W}^{\mu}(q)$ into the calculation of $\Sigma^{++\mu}(q)$. That is, we shall take
\begin{eqnarray}
	\mathcal{W}^{\mu}(q)\rightarrow \mathcal{W}^{\mu}_{\rm th}(q)=
	-2\pi \big[q^{\mu}\delta(q^2)-\hbar  B^{[\mu}n^{\nu]}q_{\nu}\delta'(q^2)\big]\tilde f(q),
\end{eqnarray}
which can be further written as 
\begin{eqnarray}\label{eq:WLB_decompose}
\mathcal{W}^{\mu}_{\rm th}(q)=
-2\pi \left[q^0n^{\mu}\delta \left(q^2+\frac{\hbar B\cdot q}{q^0}\right)-(q\cdot\hat{B})\hat{B}^{\mu}\delta \left(q^2+\frac{\hbar |B|q^0}{\hat{B}\cdot q}\right)+q^{\mu}_{\rm t}\delta(q^2)\right]\tilde f(q)
\end{eqnarray}
up to $\mathcal{O}(\hbar)$. Here, we defined $\hat{V}^{\mu}\equiv V^{\mu}/|V_{\perp}|$ and $|V_{\perp}|\equiv\sqrt{|V^2_{\perp}|}$ for an arbitrary four-vector $V^{\mu}$. Also, $q^{\mu}_{\rm t}$ satisfying $q_{\rm t}\cdot n=q_{\rm t}\cdot B=0$ denotes the transverse momentum with respect to the magnetic field. We then obtain
\begin{eqnarray}
\Sigma^{++\mu}_{\rm e}\approx -4\sqrt{2}G_{\rm F} \int \frac{\rd^4p}{(2\pi)^4} \mathcal{W}^{\mu}_{\rm th}(p)\,,
\end{eqnarray}
where $G_{\rm F} = \sqrt{2}g^2/(8M_{W}^2)$.
In the absence of magnetic fields, one immediately finds
\begin{eqnarray}
\Sigma^{++\mu}_{\rm e}|_{B=0}= \frac{\sqrt{2}}{\pi^2}G_{\rm F} n^{\mu} \int^{\infty}_0\rd|\bm p||\bm p|^2\big[\tilde f^{+}(\bm p)-\tilde f^{-}(\bm p)\big]
= \sqrt{2}G_{\rm F} N_{\rm e} n^{\mu}\,,
\end{eqnarray} 
where $\tilde f^{\pm}(\bm q)\equiv 1/[\re^{\beta(|\bm q|\mp\mu_{\rm e})}+1]$ and
\begin{align}
N_{\rm e} = \frac{\mu_{\rm e} T^2}{3} + \frac{\mu_{\rm e}^3}{3\pi^2}
\end{align}
is the electron number density. Here, we assume that the left-handed and right-handed electrons have the same chemical potential.

We can decompose the self-energy with finite magnetic fields as
\begin{eqnarray}
\Sigma^{++\mu}_{\rm e}=\mathscr{C}_0n^{\mu}+\mathscr{C}_{\rm \ell}\hat{B}^{\mu}+\mathscr{C}_{\rm t}\hat{p}^{\mu}_{\rm t}.
\end{eqnarray}
It is easy to check $\mathscr{C}_{\rm t}=0$ and
\begin{align}\nonumber
\mathscr{C}_0&= \frac{\sqrt{2}}{\pi^2}G_{\rm F} \int^{\infty}_0\rd|\bm p||\bm p|^2\int\frac{\rd\Omega}{4\pi}\left[c_{p}^+ \tilde f^{+}\left(c_{p}^+ |\bm p| \right)- c_{p}^- \tilde f^{-} \left(-c_{p}^-|\bm p| \right)\right]
\nonumber \\
&=\Sigma^{++0}|_{B=0}+\mathcal{O}(\hbar^2)\,,
\end{align}
where $c_{p}^{\pm} \equiv 1\mp {\hbar B\cdot p}/({2|\bm p|^3})$.
Here, the $B$-dependent terms vanish after the angular integration $\int \rd\Omega=\int^{2\pi}_0\rd\phi\int^{1}_{-1} \rd\cos\theta$.
On the other hand, we have a $B$-dependent contribution for $\mathscr{C}_{\rm \ell}$ as
\begin{align}\nonumber
\mathscr{C}_{\rm \ell}&=-\frac{\sqrt{2}}{\pi^2}G_{\rm F}\int^{\infty}_0\rd|\bm p||\bm p|\int\frac{\rd\Omega}{4\pi}p\cdot\hat{B}\left[\tilde f^{+}\left(|\bm p|-\frac{\hbar |B|}{2\hat{B}\cdot p}\right)
+\tilde f^{-}\left(-|\bm p|-\frac{\hbar |B|}{2\hat{B}\cdot p}\right)\right] \nonumber \\
&=\frac{\hbar G_{\rm F}}{\sqrt{2}\pi^2}|B|\int^{\infty}_0\rd|\bm p|\int\frac{\rd\Omega}{4\pi}\big[\tilde f^{+}(\bm p)-\tilde f^{-}(\bm p)\big]= \frac{\hbar G_{\rm F}}{\sqrt{2}\pi^2} \mu_{\rm e} |B|\,.
\end{align}
Since $\Sigma^{\rm r\mu}_{\rm e}=\Sigma^{++\mu}_{\rm e}-\ri\Sigma^{<\mu}_{\rm e}$ with $\Sigma^{<\mu}_{\rm e}$ being real, we have 
\begin{eqnarray}\label{eq:SigmaL_B}
	{\rm Re}(\Sigma^{\rm r\mu}_{\rm e})
	=\frac{G_{\rm F}}{\sqrt{2}} \left(N_{\rm e} n^{\mu}+ \frac{\hbar}{\pi^2} \mu_{\rm e} B^{\mu}\right)\,.
\end{eqnarray} 

To obtain the contribution from Fig.~\ref{fig:feynman}(b), we next consider the neutral-current interaction
\begin{eqnarray}
	\mathcal{L}_{\rm nc}=\frac{g}{\cos\theta_{\rm W}}\mathcal{A}^{\mu}_{ Z_0}\left[\frac{1}{2}\bar{\psi}_{\nu}\gamma_{\mu}P_{\rm L}\psi_{\nu}-\left(\frac{1}{2}-\sin^2\theta_{\rm W}\right)\bar{\psi}_{\rm e}\gamma_{\mu}P_{\rm L}\psi_{\rm e}+\sin^2\theta_{\rm W}\bar{\psi}_{\rm e}\gamma_{\mu}P_{\rm R}\psi_{\rm e}\right]\,, \nonumber \\
\end{eqnarray}
where $\mathcal{A}^{\mu}_{Z^0}$ denotes the gauge field for $Z^0$ bosons. From this interaction, we have
	\begin{eqnarray}\nonumber
		\ri\Sigma^{\delta \mu}_{\rm e}(q)&=&\frac{-g^2}{4\cos^2\theta_{\rm W}}\int \frac{\rd^4p}{(2\pi)^4}{\rm Tr}\bigg(\gamma^{\mu}\gamma^{\alpha} P_{\rm L} \bigg[\left(-\frac{1}{2}+\sin^2\theta_{\rm W}\right){\rm Tr}\big(P_{\rm L}S^{({\rm e})}(p)\gamma^{\beta}\big)
		\\
		&&+ \sin^2\theta_{\rm W}{\rm Tr}\big(P_{\rm R}S^{({\rm e})}(p)\gamma^{\beta} \big)\bigg]G^{Z}_{\alpha\beta}(q-p)\bigg)\,.
  \label{Sigma_delta}
	\end{eqnarray}
Here, $G^{Z}_{\alpha\beta}$ is the Feynman propagator for $Z^0$ bosons, which can be similarly approximated at low energy as
\begin{eqnarray}
	G^{Z}_{\alpha\beta}(q)=\frac{-\ri\eta_{\alpha\beta}}{q^2-M_{Z}^2}\approx \frac{\ri\eta_{\alpha\beta}}{M_{Z}^2}\,.
\end{eqnarray}
Taking the trace in Eq.~(\ref{Sigma_delta}) and using $M_{Z}=M_{W}/\cos\theta_{\rm W}$, we find
\begin{eqnarray}
	\ri\Sigma^{\delta \mu}_{\rm e}=\frac{-\ri g^2}{M^2_{W}}\int \frac{\rd^4p}{(2\pi)^4}\left[\left(-\frac{1}{2}+\sin^2\theta_{\rm W}\right)\mathcal{W}^{\mu}_{\rm L}(p)+\sin^2\theta_{\rm W}\mathcal{W}^{\mu}_{\rm R}(p)\right]\,.
\end{eqnarray}
Following the similar procedure for the calculation of $\ri\Sigma^{++\mu}_{\rm L, e}$, we derive
\begin{eqnarray}
	\Sigma^{\delta \mu}_{\rm e}
	=\frac{G_{\rm F}}{\sqrt{2}} \left[\left(-1+4\sin^2\theta_{\rm W}\right)N_{\rm e} n^{\mu}- \frac{\hbar}{2\pi^2} \mu B^{\mu}\right]\,.
\end{eqnarray} 

The self-energy of neutrinos in the electron background is accordingly given by
\begin{eqnarray}
	\bar{\Sigma}^{\mu}_{\rm e}={\rm Re}(\Sigma^{\rm r\mu}_{\rm e})+\Sigma^{\delta \mu}_{\rm e}=
	\frac{G_{\rm F}}{\sqrt{2}}\left[\left(1+4\sin^2\theta_{\rm W}\right) N_{\rm e} n^{\mu} + \frac{\hbar}{2\pi^2} \mu_{\rm e} B^{\mu} \right]\,.
\end{eqnarray} 
Similarly, the contributions to the neutrino self-energy from Fig.~\ref{fig:feynman}(b) in the neutron and proton backgrounds can be computed as
\begin{gather}
    \Sigma^{\delta \mu}_{\rm n}=
	-\frac{G_{\rm F}}{\sqrt{2}} N_{\rm n} n^{\mu}\,, \\
 \Sigma^{\delta \mu}_{\rm p}=
	\frac{G_{\rm F}}{\sqrt{2}}\left(1-4\sin^2\theta_{\rm W}\right) N_{\rm p} n^{\mu} \,,
\end{gather} 
respectively, where we ignored the magnetic-field contribution to $\Sigma^{\delta \mu}_{\rm p}$ that is suppressed by $1/M_{\rm p}$, with $M_{\rm p}$ being the proton mass.
Collecting altogether, the total neutrino self-energy in the electron and nucleon backgrounds is given by
\begin{eqnarray}
	\bar{\Sigma}^{\mu}=
	\frac{G_{\rm F}}{\sqrt{2}}\left(\left[\left(1+4\sin^2\theta_{\rm W}\right) N_{\rm e} - N_{\rm n} + \left(1-4\sin^2\theta_{\rm W}\right) N_{\rm p} \right] n^{\mu} + \frac{\hbar}{2\pi^2} \mu_{\rm e} B^{\mu} \right)\,.
\end{eqnarray}
The $\mathcal{O}(\hbar^0)$ contribution proportional to $n^{\mu}$, which corresponds to the potential energy $V$ in Eq.~(\ref{V}), was previously obtained in Ref.~\cite{Notzold:1987ik}. Here, we ignored the higher-order corrections of $\mathcal{O}(M_{W}^{-4})$ as in Sec.~\ref{sec:Berry}.

\subsection{Magnetic-field corrections}
We now consider the magnetic-field corrections upon the Wigner function and kinetic equation of left-handed neutrinos. To avoid complication, here we assume constant $T$ and $\mu$. From Eq.~(\ref{eq:sol_WF}), the Wigner function of left-handed neutrinos up to $\mathcal{O}(\hbar)$ and $\mathcal{O}(G_{\rm F})$ [more precisely, $\mathcal{O}(G_{\rm F} p^2)$ with $p$ being a typical energy scale of the system] is given by
\begin{eqnarray}
\label{eq:WL}
\mathcal{W}^{<\mu}(q,x) = 2\pi{\rm sgn}(q_0)\delta(\tilde{q}^2)\big(\tilde{q}^{\mu}-\hbar S_{\tilde{q}}^{\mu\nu}\mathcal{D}_{\nu}\big)f^{(\nu)}(q,x),
\end{eqnarray}
where $\tilde{q}^{\mu} = q^{\mu}-\bar{\Sigma}^{\mu}(q)$ with $\bar{\Sigma}^{\mu}(q)$ shown in Eq.~(\ref{eq:SigmaL_B}). Note here that $\mathcal{D}_{\nu}$ for neutrinos does not contain the momentum derivative coupled to electromagnetic fields.
From Eq.~(\ref{eq:offshell_CKE}), the corresponding off-shell kinetic equation reads
\begin{eqnarray}\label{eq:CKE_B}
\tilde q^{\mu} \partial_{\mu}f^{(\nu)}(q,x)=C[f^{(\nu)}].
\end{eqnarray}
The explicit form of $C[f^{(\nu)}]$ incorporating the magnetic-field corrections can be found in Ref.~\cite{Yamamoto:2020zrs}, which is not of our interest in the present work. 
Note that the terms proportional to $\delta'(q^2)$ due to the spacetime derivative upon magnetic fields appearing from $\partial_{\mu}\mathcal{W}^{<\mu}$ and from $(\partial^{\nu}_{q}\mathcal{W}^{<\mu})(\partial_{\nu}\bar{\Sigma}^{\mu})$ cancel each other. Alternatively, the on-shell kinetic equation is given by
\begin{eqnarray}
\Big[(|\bm q|+\Delta \epsilon_{qB})\partial_0+(q_{\perp}^{\mu}-\bar{\Sigma}^{\mu}_B)\partial_{\mu}\Big]f^{(\nu)}({\bm q},x)=(|\bm q|n^{\nu}+q_{\perp}^{\nu})C_{\nu}[f^{(\nu)}]\,,
\end{eqnarray}
where $\Delta \epsilon_q=\hat{q}\cdot\bar{\Sigma}$ and $\Delta \epsilon_{qB}=\hat{q}\cdot\bar{\Sigma}_B$ with $\bar{\Sigma}^{\mu}_B=\hbar G_{\rm F} \mu_{\rm e} B^{\mu}/(2\sqrt2 \pi^2)$.
When $B^{\mu}=0$, this reduces to the conventional Boltzmann equation. For practical purposes, it is more convenient to rewrite Eq.~(\ref{eq:WL}) as 
\begin{eqnarray}\label{eq:WL_practical}
\mathcal{W}^{<\mu}(q,x)\approx \pi \frac{\delta(q_0-|\bm q|)}{|{\bm q}|}
\Big[|\bm q|n^{\mu}+\left(|\bm q|-\Delta\epsilon_{q}\right)(\hat{q}_{\perp}^{\mu}-\hbar S^{\mu\nu}_q\partial_{\nu})-\bar{\Sigma}^{\mu}\Big]
\Big(f^{(\nu)}+\Delta\epsilon_{q}\partial_{q_0}f^{(\nu)}\Big)
\nonumber \\
\end{eqnarray}
for $f^{(\nu)}=f^{(\nu)}(q,x)$ with $q_0>0$ up to $\mathcal{O}(\hbar)$ and $\mathcal{O}(G_{\rm F})$.

Assuming $f^{(\nu)}(q,x)=\Theta(q_0)f^{(\nu)}(q)$, we may evaluate the neutrino number current induced by magnetic fields as
\begin{align}
\label{eq:eff_CME}
\nonumber
J^{\mu}_{B}&=2\int\frac{\rd^4q}{(2\pi)^4}\left[\mathcal{W}^{<\mu}(q,x)-\mathcal{W}^{<\mu}(q,x)_{B=0}\right]
\\\nonumber
&=\int\frac{\rd^4q}{(2\pi)^3}\frac{\delta(q_0-|\bm q|)}{q_0}\Theta(q_0)\big[\Delta \epsilon_{qB}\big(q^{\mu}\partial_{q^0}-\hat{q}^{\mu}_{\perp}\big)-\bar{\Sigma}^{\mu}_{B}\big]f^{(\nu)}(q)
\\
&= \frac{\hbar G_{\rm F}}{\sqrt{2}\pi^2} \mu_{\rm e}|B|\int\frac{\rd^4q}{(2\pi)^3}\frac{\delta(q_0-|\bm q|)}{2q_0}\Theta(q_0)\left[\hat{q}^{\mu}\hat{q}\cdot\hat{B}(q_0\partial_{q_0}-1)+n^{\mu} \hat q \cdot B -\hat{B}^{\mu}\right] f^{(\nu)}(q)\,.
\end{align}
One can show by the integration by parts that this current vanishes when neutrinos are also in thermal equilibrium, $f^{(\nu)}(q)=\bar f^{(\nu)\pm}(q_0)$, where $\bar f^{(\nu)\pm}(q_0)=1/[{\rm e}^{\beta (q_0 \mp \mu_{\nu})}+1]$ is the equilibrium distribution function for neutrinos and antineutrinos, respectively, with $\mu_{\nu}$ being the neutrino chemical potential. This result is consistent with the generalized Bloch theorem~\cite{Yamamoto:2015fxa} that any particle number current should vanish in thermal equilibrium. On the other hand, the neutrino number and energy currents may receive corrections from magnetic fields for an anisotropic neutrino distribution. 

In the context of core-collapse supernovae, there are several mechanisms for the anisotropy of the neutrino distribution, such as the strong magnetic field~\cite{Vilenkin:1995um,Yamamoto:2021hjs}, rotation~\cite{Takiwaki:2017tpe}, convection in the postshock layer and the neutrino sphere~\cite{Wongwathanarat:2012zp}, and so on. However, since it has not yet been established which mechanism is most effective, in this study, we assume the presence of anisotropy without relying on a specific mechanism and provide a proof-of-principle demonstration for this novel chiral transport. To investigate the relevance of this effect quantitatively, three-dimensional numerical simulations of the chiral radiation hydrodynamics would be necessary, which is beyond the scope of the present paper.

As a demonstration, one may consider an anisotropic distribution function with the form
\begin{eqnarray}\label{eq:fL_assumpt2}
f^{(\nu)}(q)=f \left(\sqrt{q^{\mu}\Xi_{\mu\nu}q^{\nu}} \right)
\end{eqnarray}
with $\Xi_{\mu\nu}=\bar{n}_{\mu}\bar{n}_{\nu}+\xi \hat{a}_{\mu}\hat{a}_{\nu}$, where $\hat{a}^{\mu}$ corresponds to a unit spacelike vector and $\xi$ represents the magnitude of anisotropy. Such a distribution function is introduced in anisotropic hydrodynamics as a model for the early-time dynamics with large pressure anisotropy in relativistic heavy ion collisions; see, e.g., Ref.~\cite{Strickland:2014pga} for a review. When $|\xi|\ll 1$, we may also make the approximation
\begin{align}
f \left(\sqrt{q^{\mu}\Xi_{\mu\nu}q^{\nu}} \right)=f \left(\sqrt{q_0^2+\xi(q_{\perp}\cdot \hat{a})^2} \right)\approx f(q_0)+ \frac{\xi(q_{\perp}\cdot \hat{a})^2}{2q_0}\partial_{q_0}f(q_0)\,, 
\end{align}
where the leading-order anisotropic correction is given by the quadratic term with respect to the anisotropic vector $\hat{a}^{\mu}$. 
For simplicity, we further assume that $\hat{a}^{\mu}$ is either perpendicular or parallel to $B^{\mu}$.
Accordingly, Eq.~(\ref{eq:eff_CME}) yields
\begin{align}\nonumber\label{eq:CME_current}
J^{\mu}_{B}&\approx -\frac{\hbar G_{\rm F}}{\sqrt{2}\pi^2}\mu_{\rm e}|B|\int\frac{\rd^4q}{(2\pi)^3}\frac{\delta(q_0-|\bm q|)}{4q_0^2}\Theta(q_0)\xi(q_{\perp}\cdot\hat{a})^2
\left[\hat{B}^{\mu}+\hat{q}^{\mu}\hat{q}\cdot\hat{B}(2-q_0\partial_{q_0})\right]\partial_{q_0}f(q_0)
\\
&=-\frac{\hbar G_{\rm F}}{\sqrt{2}\pi^2} \mu_{\rm e}|B|\int\frac{\rd^3q}{(2\pi)^3}\frac{\xi(q_{\perp}\cdot\hat{a})^2}{4|\bm q|^2}
\left(\hat{B}^{\mu}+5\hat{q}^{\mu}\hat{q}\cdot\hat{B}\right)\partial_{|\bm q|}f(|\bm q|)
\nonumber \\
&=\frac{\hbar G_{\rm F}}{12\sqrt{2}\pi^4} \mu_{\rm e} \xi (\hat{B}\cdot\hat{a})^2 B^{\mu} \int^{\infty}_0 \rd|\bm q||\bm q|^2\partial_{|\bm q|}f(|\bm q|)\,.
\end{align}
In this example, the neutrino number current can be generated along the magnetic field.

We may also calculate the modifications upon the (symmetric) energy-momentum tensor by using
\begin{align}\nonumber
T^{\mu\nu}_{B}&=\int\frac{\rd^4q}{(2\pi)^4}q^{(\nu}\left[\mathcal{W}^{<\mu)}(q,x)-\mathcal{W}^{<\mu)}(q,x)_{B=0}\right]
\\
&=\frac{1}{2}\int\frac{\rd^4q}{(2\pi)^3}\delta(q_0-|\bm q|)\Theta(q_0)
\Big[
\Delta\epsilon_{qB} \left(n^{(\mu}n^{\nu)} + n^{(\mu} \hat q_{\perp}^{\nu)} \right) q_0\partial_{q_0} 
+ \Delta\epsilon_{qB} \hat{q}_{\perp}^{(\mu}\hat{q}_{\perp}^{\nu)}
\big(q_0\partial_{q_0}-1\big) \nonumber\\
& \qquad \quad - \hat{q}^{(\mu}\bar{\Sigma}_B^{\nu)}\Big]f^{(\nu)}(q)\,.
\label{eq:T_B}
\end{align}
By symmetry, we can decompose $T^{\mu\nu}_{B}$ as%
\footnote{One may alternatively rewrite $\xi_{\rm t}^{(\mu}\xi_{\rm t}^{\nu)}$ as $\xi_{\rm t}^{(\mu}\xi_{\rm t}^{\nu)}=-2(\eta^{\mu\nu}-n^{(\mu}n^{\nu)}+\hat{B}^{\mu}\hat{B}^{\nu})$.} 
\begin{eqnarray}
T^{\mu\nu}_B=\chi_{B1}n^{(\mu}n^{\nu)}+\chi_{B2}n^{(\mu}\hat{B}^{\nu)}+\chi_{B3}\hat{B}^{(\mu}\hat{B}^{\nu)}+\chi_{B4}\xi_{\rm t}^{(\mu}\hat{B}^{\nu)}+\chi_{B5}\xi_{\rm t}^{(\mu}\xi_{\rm t}^{\nu)},
\end{eqnarray}
where $\xi_{\rm t}^{\mu}$ is a unit spacelike vector satisfying $\xi_{\rm t}\cdot n=\xi_{\rm t}\cdot \hat{B}=0$. The corresponding coefficients are given by
\begin{align}
\chi_{B1}&=\frac{\hbar G_{\rm F}}{4 \sqrt2 \pi^2} \mu_{\rm e}|B|\int\frac{\rd^4q}{(2\pi)^3}\delta(q_0-|\bm q|)\Theta(q_0)(\hat{q}\cdot\hat{B})q_0\partial_{q_0}f^{(\nu)}(q)\,,
\\
\chi_{B2}&=-\frac{\hbar G_{\rm F}}{4 \sqrt2 \pi^2} \mu_{\rm e}|B|\int\frac{\rd^4q}{(2\pi)^3}\delta(q_0-|\bm q|)\Theta(q_0)\left[1+(\hat{q}\cdot \hat{B})^2 q_0\partial_{q_0} \right]f^{(\nu)}(q)\,,
\\
\chi_{B3}&=\frac{\hbar G_{\rm F}}{4 \sqrt2 \pi^2} \mu_{\rm e}|B|\int\frac{\rd^4q}{(2\pi)^3}\delta(q_0-|\bm q|)\Theta(q_0)(\hat{q}\cdot\hat{B})\left[1+(\hat{q}\cdot \hat{B})^2 \big(q_0\partial_{q_0}-1 \big)\right]f^{(\nu)}(q)\,,
\\
\chi_{B4}&=\frac{\hbar G_{\rm F}}{4 \sqrt2 \pi^2} \mu_{\rm e}|B|\int\frac{\rd^4q}{(2\pi)^3}\delta(q_0-|\bm q|)\Theta(q_0)(\xi_{\rm t}\cdot\hat{q})\left[2(\hat{q}\cdot\hat{B})\big(q_0\partial_{q_0}-1\big)-1\right]f^{(\nu)}(q)\,,
\\
\chi_{B5}&=\frac{\hbar G_{\rm F}}{4 \sqrt2 \pi^2} \mu_{\rm e}|B|\int\frac{\rd^4q}{(2\pi)^3}\delta(q_0-|\bm q|)\Theta(q_0)(\xi_{\rm t}\cdot\hat{q})^2(\hat{q}\cdot\hat{B})\big(q_0\partial_{q_0}-1\big)f^{(\nu)}(q)\,.
\end{align}
It is also clear to see that all the coefficients above vanish when $f^{(\nu)}(q)=\bar f^{(\nu)\pm}(q_0)$.

\subsection{Neutrino spin Hall effect}
We next analyze the neutrino spin Hall effect induced by the density gradient. For clarity, we shall now switch off the magnetic field. Given Eqs.~(\ref{eq:SigmaL_B}) and (\ref{eq:sol_WF}), we find   
\begin{eqnarray}
	\mathcal{W}^{<\mu}=2\pi {\rm sgn}(q_0)\Big[\delta(\tilde{q}^2)\big(\tilde{q}^{\mu}-\hbar S^{(\nu)\mu\nu}_{q}\mathcal{D}_{\nu}\big)
	-\hbar \delta'(q^2)\epsilon^{\mu\nu\rho\sigma}q_{\nu}n_{\sigma}(\partial_{\rho}V)\Big]f^{(\nu)}\,,
\end{eqnarray}
where we used the decomposition $\bar{\Sigma}_{\mu}=V n_{\mu}$, with $V$ given in Eq.~(\ref{V}). 
On the other hand, the off-shell chiral kinetic equation becomes
\begin{eqnarray}\label{eq:CKE_Tmu}
	\left[\tilde{q}^{\mu}\left(\partial_{\mu}+n_{\mu}(\partial_{\nu}V)\partial^{\nu}_{q}\right)-\frac{\hbar S^{\mu\nu}_q}{q\cdot n}(\partial_{\mu}V)\partial_{\nu}
	\right]f^{(\nu)} = C[f^{(\nu)}]\,.
\end{eqnarray}

Similar to the case with magnetic fields, the neutrino number current due to the term $\partial_{\rho}V$ is given by
\begin{align}
\label{eq:J_SHE}
J^{\mu}_{\rm SHE}&=2\int\frac{\rd^4q}{(2\pi)^4}\left[\mathcal{W}^{<\mu}(q,x)-\mathcal{W}^{<\mu}(q,x)_{T,\mu={\rm const.}}\right]
\nonumber\\
&=\hbar \epsilon^{\mu\nu\rho\sigma} \ell_{\nu} n_{\sigma}(\partial_{\rho}V)\,,
\end{align}
where
\begin{equation}
\label{eq:ell}
\ell_{\nu} = \int\frac{\rd^4q}{(2\pi)^3}\Theta(q_0)\frac{\delta(q_0-|\bm q|)}{2q_0}\partial_{q\nu}f^{(\nu)}(q)\,.
\end{equation}
This spatial ($\mu=i$) part exactly reproduces the neutrino spin Hall effect in Eq.~(\ref{eq:SHE}) derived using Berry curvature in Sec.~\ref{sec:Berry}. Hence, we can confirm that this kinetic theory based on the Wigner function correctly provides a Lorentz-covariant formulation including the effects of Berry curvature.

As we already remarked in Sec.~\ref{sec:Berry}, $J^{\mu}_{\rm SHE}$ can be nonvanishing only when the momentum distribution of neutrinos is anisotropic. As an example, we take 
\begin{eqnarray}\label{eq:fL_assumpt}
f^{(\nu)}(q)=f(q\cdot v)\approx f(q_0) + (q_{\perp} \cdot v) \partial_{q_0}f(q_0),
\end{eqnarray}
where the vector $v^{\mu}=(1,{\bm v})$ with $|{\bm v}|\ll1$ characterizes an anisotropy of the neutrino distribution function. Note that Eq.~(\ref{eq:fL_assumpt}) satisfies Eq.~(\ref{eq:CKE_Tmu}) without collisions given $v\cdot\partial V=0$. In this case, we find
\begin{align}
\ell_{\nu}&\approx -\frac{1}{24\pi^2}v_{\nu\perp}\int^{\infty}_0 \rd|\bm q||\bm q|\partial_{|\bm q|}f(|\bm q|)\,.
\end{align}  
On the contrary, Eq.~(\ref{eq:fL_assumpt2}) leads to $\ell_{\nu}=0$ and $J^{\mu}_{\rm SHE}=0$. 

One may further evaluate the modifications on the energy-momentum tensor via
\begin{align}
	T^{\mu\nu}_{\rm SHE}&=\int\frac{\rd^4q}{(2\pi)^4}q^{(\nu}\left[\mathcal{W}^{<\mu)}(q,x)-\mathcal{W}^{<\mu)}(q,x)_{T,\mu={\rm const.}}\right]
	\nonumber\\
	&=\epsilon^{(\mu\kappa\rho\sigma}L^{\nu)}_{\kappa}n_{\sigma}(\partial_{\rho}V)\,,
 \label{eq:T_SHE}
\end{align}
where
\begin{align}
L^{\nu}_{\kappa} = \frac{1}{2}\int\frac{\rd^4q}{(2\pi)^3}\Theta(q_0)\frac{\delta(q_0-|\bm q|)}{2q_0}q^{\nu}\partial_{q\kappa}f^{(\nu)}(q)\,.
\end{align}
We similarly find that $T^{\mu\nu}_{\rm SHE}=0$ when $f^{(\nu)}(q)=\bar f^{(\nu)\pm}(q_0)$.

\section{Summary and outlook}\label{sec:summary}
In this paper, we derived the CKT for chiral fermions with self-energy corrections. By applying this formalism to nonequilibrium neutrinos interacting with electrons in equilibrium, we found the neutrino currents along the magnetic fields and neutrino spin Hall effect induced by the density gradient. Combined with the quantum corrections to the collision terms found previously~\cite{Yamamoto:2020zrs}, the present results provide a more complete CKT.

While the magnetic-field-induced neutrino current previously found in Ref.~\cite{Yamamoto:2021hjs} is $\mathcal{O}(G_{\rm F}^2)$, the one found in this paper is $\mathcal{O}(G_{\rm F})$, and the latter is naively more dominant in $G_{\rm F}$. On the other hand, the latter requires the anisotropic neutrino distribution, which may be generated by a strong magnetic field~\cite{Vilenkin:1995um,Yamamoto:2021hjs}, rotation~\cite{Takiwaki:2017tpe}, convective motion~\cite{Wongwathanarat:2012zp}, etc., in core-collapse supernovae. Which contribution is more effective is a dynamical question that needs to be checked by numerical simulations of chiral radiation hydrodynamics. This issue would be important for pulsar kicks, to which these neutrino chiral transport phenomena contribute; see Ref.~\cite{Kamada:2022nyt} and references therein.

Our results indicate the potential relevance of the neutrino spin Hall effect in the physics of core-collapse supernovae. Similarly, the electron spin Hall effect due to the neutrino density gradient should also appear at the core of supernovae, where neutrinos are thermalized. To our knowledge, this is the first example of the spin Hall effect of fermions realized in nature. It would be interesting to investigate how the presence of these spin Hall effects modifies the evolution of core-collapse supernovae. While we focused on the flat spacetime in this paper, extensions to curved spacetime would lead to an additional spin Hall effect of leptons due to the gravitational field similar to photons~\cite{Gosselin:2006wp,Oancea:2020khc,Mameda:2022ojk} and gravitons~\cite{Yamamoto:2017gla,Andersson:2020gsj}. Incorporating these chiral transport phenomena, which are inevitable consequences of the Standard Model, would provide genuine first-principles simulations of supernovae.

A further extension to the quantum kinetic theory for massive fermions \cite{Hattori:2019ahi,Yang:2020hri}, similarly with the additional self-energy corrections, has been recently reported in Ref.~\cite{Fang:2023bbw}, in which the application to spin polarization phenomena in relativistic heavy ion collisions is also addressed.

\acknowledgments
N.~Y.~is supported in part by the Keio Institute of Pure and Applied Sciences (KiPAS) project at Keio University and JSPS KAKENHI Grant No. JP19K03852 and No. JP22H01216. 
D.-L. Y.~is supported by National Science and Technology Council (Taiwan) under Grant No. MOST 110-2112-M-001-070-MY3. The Feynman diagrams in this paper are generated using \texttt{TikZ-Feynman}~\cite{Ellis:2016jkw}.

\appendix

\section{Perturbative solutions of the master equations}\label{app:KBeq_solution}

In this appendix, we provide the details of the perturbation solutions of the master equations (\ref{eq:master_onshellcond}) and (\ref{eq:master_auxiliary}). We will focus on the case for left-handed fermions without loss of generality and drop the contributions of the terms $\mathcal{O}(\bar{\Sigma}_{\rm L}^2)$, $\mathcal{O}(\Sigma^{\lessgtr}_{\rm L}\bar{\Sigma}_{\rm L})$, and ${\cal O}\big((\Sigma^{\lessgtr}_{\rm L})^2 \big)$ at weak coupling. In the following, we will frequently use the useful relations
\begin{gather}
\label{Delta_q}
    \tilde{\Delta}_{\rho}\tilde{q}_{\sigma}\approx \Delta_{[\sigma}\bar{\Sigma}_{\rm L\rho]}+F_{\sigma\rho}, \\
\label{relations}
(\tilde{q}\cdot\tilde{\Delta})\tilde{q}^2\approx 0, \quad \tilde{\Delta}\cdot\tilde{q}\approx 0, \quad 
\tilde{q}\cdot\tilde{\Delta}\tilde{q}_{\sigma}\approx -\tilde{q}^{\alpha}\tilde{\Delta}_{\sigma}\tilde{q}_{\alpha}, \quad 
\epsilon^{\mu\nu\rho\sigma}(\tilde{\Delta}_{\mu}\tilde{\Delta}_{\sigma}\tilde{q}_{\rho}) \approx 0,
\end{gather}
and the Schouten identity
\begin{eqnarray}
\label{Schouten}
q^{\alpha}\epsilon^{\mu\nu\rho\sigma}=q^{\mu}\epsilon^{\alpha\nu\rho\sigma}+q^{\nu}\epsilon^{\mu\alpha\rho\sigma}+q^{\rho}\epsilon^{\mu\nu\alpha\sigma}+q^{\sigma}\epsilon^{\mu\nu\rho\alpha}.
\end{eqnarray}

To find a perturbative solution, we decompose $\mathcal{W}^{<\mu}_{\rm L}=\mathcal{W}^{(0)<\mu}_{\rm L}+\hbar \mathcal{W}^{(1)<\mu}_{\rm L}$, where $\mathcal{W}^{(0)<\mu}_{\rm L}$ is given by Eq.~(\ref{eq:solution_leading}) for $\chi={\rm L}$. 
Given
\begin{eqnarray}
	\Delta_{\mu}\mathcal{W}^{(0)<\mu}_{\rm L}\approx 2\pi{\rm sgn}(q_0)\left[\delta(\tilde{q}^2)\big(\tilde{q}\cdot\Delta- \Delta\cdot\bar{\Sigma}_{\rm L}\big)-2\delta'({\tilde q}^2)\tilde{q}^{\mu}(\tilde{q}\cdot\Delta \bar{\Sigma}_{\rm L\mu})\right]f_{\rm L}
\end{eqnarray}
and
\begin{align}\nonumber
	&(\partial^{\nu}_{q}\mathcal{W}^{(0)<\mu}_{\rm L})(\Delta_{\nu}\bar{\Sigma}_{\rm L\mu})-(\partial_{\nu}\mathcal{W}^{(0)<\mu}_{\rm L})(\partial^{\nu}_{q}\bar{\Sigma}_{\rm L\mu})
	\\
	&\approx 2\pi {\rm sgn}(q_0)\Big(2\delta'(\tilde{q}^2)\tilde{q}^{\mu}(\tilde{q}\cdot\Delta \bar{\Sigma}_{\rm L\mu})+\delta(\tilde{q}^2) \big[\Delta\cdot\bar{\Sigma}_{\rm L}
	+\tilde{q}^{\mu}(\Delta_{\nu}\bar{\Sigma}_{\rm L\mu})\partial^{\nu}_{q}
	-\tilde{q}^{\mu}(\partial^{\nu}_{q}\bar{\Sigma}_{\rm L\mu})\partial_{\nu} \big]
	\Big)f_{\rm L}\,,
\end{align}
we obtain
\begin{eqnarray}
\tilde{\Delta}_{\mu}\mathcal{W}^{(0)<\mu}_{\rm L}-\Sigma^{<}_{\rm L\mu}\mathcal{W}^{(0)>\mu}_{\rm L}+\Sigma^{>}_{\rm L\mu}\mathcal{W}^{(0)<\mu}_{\rm L}=\delta(\tilde{q}^2)\tilde{q}^{\mu}\big(\tilde{\Delta}_{\mu}f_{\rm L}-\mathcal{C}_{\mu}[f_{\rm L}]\big)
=\mathcal{O}(\hbar),
\end{eqnarray}
which leads to the kinetic equation~(\ref{KE_leading}) at $\mathcal{O}(\hbar^0)$, where $\mathcal{C}_{\mu}[f_{\rm L}]$ and $\tilde{\Delta}_{\rho}$ are defined below Eq.~(\ref{KE_leading}).  

To obtain $\mathcal{W}^{(1)<\mu}_{\rm L}$, we have to solve Eqs.~(\ref{eq:W1_q}) and (\ref{eq:W1_n}) with the constraint $\tilde{q}_{\mu}\mathcal{W}^{(1)<\mu}_{\chi}=0$.
Inserting
\begin{eqnarray}
-\frac{1}{2}\epsilon^{\mu\nu\rho\sigma}\tilde{\Delta}_{\rho}\mathcal{W}^{(0)<}_{\rm L\sigma}
\approx-\pi {\rm sgn}(q_0)\epsilon^{\mu\nu\rho\sigma}\left[\delta(\tilde{q}^2)\tilde{\Delta}_{\rho}(\tilde{q}_{\sigma}f_{\rm L})+2\delta'(\tilde{q}^2)f_{\rm L}\tilde{q}_{\sigma}\tilde{q}^{\alpha}(\Delta_{[\alpha}\bar{\Sigma}_{\rho]}+F_{\alpha\rho})\right]
\nonumber \\
\end{eqnarray}
into Eq.~(\ref{eq:W1_q}) yields
\begin{eqnarray}
\tilde{q}^{2}\mathcal{W}^{(1)<\mu}_{\rm L}
	=-\pi {\rm sgn}(q_0)\epsilon^{\mu\nu\rho\sigma}\tilde{q}_{\nu}(\tilde{\Delta}_{\rho}\tilde{q}_{\sigma})\delta(\tilde{q}^2)f_{\rm L}\,,
\end{eqnarray}
from which we may postulate the solution (\ref{eq:W1_solution}).

For generality, here we consider a spacetime-dependent frame vector $n^{\mu}(x)$. By using the Schouten identity (\ref{Schouten}) and the leading-order kinetic equation, one can show that
\begin{align}\nonumber
\tilde{q}^{[\nu}\mathcal{W}^{(1)<\mu]}_{\rm L}&=
-2\pi {\rm sgn}(q_0)(\tilde{q}^{\kappa}\epsilon^{\mu\nu\rho\sigma}+\tilde{q}^{\rho}\epsilon^{\mu\kappa\nu\sigma}+\tilde{q}^{\sigma}\epsilon^{\mu\kappa\rho\nu})\bigg[\delta(\tilde{q}^2)\frac{\tilde{q}_{\sigma}n_{\kappa}}{2\tilde{q}\cdot n}\big(\tilde{\Delta}_{\rho}f_{\rm L}-\mathcal{C}_{\rho}[f_{\rm L}]\big)
\\\nonumber
&\quad-\frac{1}{2}\delta'(\tilde{q}^2)\tilde{q}_{\kappa}(\tilde{\Delta}_{\rho}\tilde{q}_{\sigma})f_{\rm L}\bigg]
\\\nonumber
&\approx-\pi {\rm sgn}(q_0)\epsilon^{\mu\nu\rho\sigma}\left(\delta(\tilde{q}^2)\big[\tilde{\Delta}_{\rho}(\tilde{q}_{\sigma}f_{\rm L})-\tilde{q}_{\sigma}\mathcal{C}_{\rho}[f_{\rm L}]\big]+2\delta'(\tilde{q}^2)f_{\rm L}\tilde{q}_{\sigma}\tilde{q}^{\alpha}(\Delta_{[\alpha}\bar{\Sigma}_{\rho]}+F_{\alpha\rho})\right)
\\
&=-\frac{1}{2}\epsilon^{\mu\nu\rho\sigma}\big(\tilde{\Delta}_{\rho}\mathcal{W}^{(0)<}_{\rm L\sigma}-\Sigma^{<}_{\rm L\rho}\mathcal{W}^{(0)>}_{\rm L\sigma}+\Sigma^{>}_{\rm L\rho}\mathcal{W}^{(0)<}_{\rm L\sigma}\big)
\end{align} 
and $\tilde{q}_{\mu}\mathcal{W}^{(1)<\mu}_{\rm L}=0$. Therefore, Eq.~(\ref{eq:W1_solution}) is indeed the perturbative solution satisfying the Kadanoff-Baym equations. One may analogously derive the perturbative solution for right-handed fermions. 

From Eq.~(\ref{eq:master_KE}) and the perturbative solution of $\mathcal{W}^{\mu}_{\rm L}$ in Eq.~(\ref{eq:sol_WF}), the corresponding chiral kinetic equation reads
\begin{align}\nonumber\label{eq:CKE_L_1}
&\bigg(\delta(\tilde{q}^2)\big(\tilde{q}^{\mu}\tilde{\Delta}_{\mu}-\hbar \tilde{\Delta}_{\mu}S^{\mu\nu}_{\tilde{q}}\tilde{\Delta}_{\nu}\big)
-\hbar\delta'(\tilde{q}^2)\left[(\tilde{\Delta}_{\mu}\tilde{q}^2)S^{\mu\nu}_{\tilde{q}}\tilde{\Delta}_{\nu}
+\frac{1}{2}\epsilon^{\mu\nu\rho\sigma}\tilde{\Delta}_{\mu}\tilde{q}_{\nu}(\tilde{\Delta}_{\sigma}\tilde{q}_{\rho})\right]
\\
&-\frac{\hbar}{2}\epsilon^{\mu\nu\rho\sigma}\delta''(\tilde{q}^2)(\tilde{\Delta}_{\mu}\tilde{q}^2)\tilde{q}_{\nu}(\tilde{\Delta}_{\sigma}\tilde{q}_{\rho})\bigg)f_{\rm L}=\tilde{C}[f_{\rm L}]+\hbar \delta'(\tilde{q}^2)S_{\tilde{q}}^{\mu\nu}F_{\mu\nu} \tilde{q}\cdot \mathcal{C}[f_{\rm L}]\,,
\end{align}
where we used the Schouten identity (\ref{Schouten}). Here, $\tilde{C}[f_{\rm L}]$ is the collision term involving $\Sigma^{\lessgtr}_{\rm L}$, which, dropping the $\mathcal{O}(\Sigma^{\lessgtr}_{\rm L}\bar{\Sigma}_{\rm L})$ terms, takes the form $\tilde{C}[f_{\rm L}]=\delta(\tilde{q}^2)C[f_{\rm L}]$ with $C[f_{\rm L}]$ given in Eq.~(\ref{eq:C})~\cite{Kamada:2022nyt}.

We can rewrite this kinetic equation such that the terms proportional to  $\delta''(\tilde{q}^2)$ and $\delta'(\tilde{q}^2)$ do not appear explicitly. By using the Schouten identity (\ref{Schouten}), we have
\begin{align}
\frac{1}{2}\epsilon^{\mu\nu\rho\sigma}(\tilde{\Delta}_{\mu}\tilde{q}^2)\tilde{q}_{\nu}(\tilde{\Delta}_{\sigma}\tilde{q}_{\rho})&=\epsilon^{\mu\nu\rho\sigma}\tilde{q}^{\alpha}(\tilde{\Delta}_{\mu}\tilde{q}_{\alpha})\tilde{q}_{\nu}(\tilde{\Delta}_{\sigma}\tilde{q}_{\rho})
\nonumber \\
&=-\epsilon^{\mu\nu\rho\sigma}\big[\tilde{q}^{\alpha}(\tilde{\Delta}_{\mu}\tilde{q}_{\alpha})\tilde{q}_{\nu}-\tilde{q}^2(\tilde{\Delta}_{\mu}\tilde{q}_{\nu})
-2\tilde{q}_{\nu}(\tilde{q}\cdot\tilde{\Delta}\tilde{q}_{\mu})\big](\tilde{\Delta}_{\sigma}\tilde{q}_{\rho}),
\end{align}
which yields
\begin{eqnarray}
\frac{1}{2}\epsilon^{\mu\nu\rho\sigma}\delta''(\tilde{q}^2)(\tilde{\Delta}_{\mu}\tilde{q}^2)\tilde{q}_{\nu}(\tilde{\Delta}_{\sigma}\tilde{q}_{\rho})
=\frac{1}{4}\epsilon^{\mu\nu\rho\sigma}\tilde{q}^2\delta''(\tilde{q}^2)(\tilde{\Delta}_{\mu}\tilde{q}_{\nu})(\tilde{\Delta}_{\sigma}\tilde{q}_{\rho})
=-\frac{1}{2}\epsilon^{\mu\nu\rho\sigma}\delta'(\tilde{q}^2)(\tilde{\Delta}_{\mu}\tilde{q}_{\nu})(\tilde{\Delta}_{\sigma}\tilde{q}_{\rho})\,.
\nonumber \\
\end{eqnarray}
One can similarly show that
\begin{eqnarray}
(\tilde{\Delta}_{\mu}\tilde{q}^2)S^{\mu\nu}_{\tilde{q}}\tilde{\Delta}_{\nu}=\frac{1}{2}\epsilon^{\mu\nu\rho\sigma}(\tilde{\Delta}_{\sigma}\tilde{q}_{\rho})\left(\frac{\tilde{q}^2n_{\nu}}{\tilde{q}\cdot n}-\tilde{q}_{\nu}\right)\tilde{\Delta}_{\mu}+S^{\mu\nu}_{\tilde{q}}(\tilde{\Delta}_{\mu}\tilde{q}_{\nu})\tilde{q}\cdot\tilde{\Delta}\,.
\end{eqnarray}
Using these relations, Eq.~(\ref{eq:CKE_L_1}) becomes
\begin{align}\label{eq:CKE_L_3}
	&\left(\delta(\tilde{q}^2)\left[\tilde{q}^{\mu}\tilde{\Delta}_{\mu}-\hbar \tilde{\Delta}_{\mu}S^{\mu\nu}_{\tilde{q}}\tilde{\Delta}_{\nu}-\frac{\hbar\epsilon^{\mu\nu\rho\sigma}n_{\mu}(\tilde{\Delta}_{\sigma}\tilde{q}_{\rho})}{2\tilde{q}\cdot n}\tilde{\Delta}_{\nu}\right]
	-\frac{\hbar}{2}\delta'(\tilde{q}^2)\epsilon^{\mu\nu\rho\sigma}\tilde{q}_{\nu}(\tilde{\Delta}_{\mu}\tilde{\Delta}_{\sigma}\tilde{q}_{\rho})
	\right)f_{\rm L}
	 \nonumber \\
	&\quad -\hbar \delta'(\tilde{q}^2)S^{\alpha\beta}_{\tilde{q}}(\tilde{\Delta}_{\alpha}\tilde{q}_{\beta})\tilde{q}^{\mu} \big(\tilde{\Delta}_{\mu}f_{\rm L}-\mathcal{C}_{\mu}[f_{\rm L}]\big)=\tilde{C}[f_{\rm L}]\,.
\end{align}
We may eliminate the term proportional to $\delta'(\tilde{q}^2)$ in the first line of Eq.~(\ref{eq:CKE_L_3}) from Eq.~(\ref{relations}). Then, Eq.~(\ref{eq:CKE_L_3}) can be written as
\begin{eqnarray}
	\delta\left[\tilde{q}^2-\hbar S^{\alpha\beta}_{\tilde{q}}(\tilde{\Delta}_{\alpha}\tilde{q}_{\beta})\right]\left( \left[\tilde{q}^{\mu}\tilde{\Delta}_{\mu}-\hbar \tilde{\Delta}_{\mu}S^{\mu\nu}_{\tilde{q}}\tilde{\Delta}_{\nu}-\frac{\hbar\epsilon^{\mu\nu\rho\sigma}n_{\mu}(\tilde{\Delta}_{\sigma}\tilde{q}_{\rho})}{2\tilde{q}\cdot n}\tilde{\Delta}_{\nu}\right]f_{\rm L} -C[f_{\rm L}]\right)=0\,,
	\nonumber \\
\end{eqnarray}
up to $\mathcal{O}(\hbar)$. By further using
\begin{align}\nonumber
	-\tilde{\Delta}_{\mu}S^{\mu\nu}_{\tilde{q}}\tilde{\Delta}_{\nu}&=\frac{S^{\mu\nu}_{\tilde{q}}}{\tilde{q}\cdot n} \left(\big[(\Delta_{[\alpha}\bar{\Sigma}_{\rm L\mu]})+F_{\alpha\mu}\big]n^{\alpha} + \tilde{q}^{\alpha}(\tilde{\Delta}_{\mu}n_{\alpha})\right)\tilde{\Delta}_{\nu}
	\\
	& \quad +\frac{\epsilon^{\mu\nu\rho\sigma}}{2\tilde{q}\cdot n}\left[n_{\mu}(\tilde{\Delta}_{\sigma}\tilde{q}_{\rho})+\tilde{q}_{\rho}(\tilde{\Delta}_{\sigma}n_{\mu})\right]\tilde{\Delta}_{\nu} -S^{\mu\nu}_{\tilde{q}}\tilde{\Delta}_{\mu}\tilde{\Delta}_{\nu}\,,
\end{align}
we eventually arrive at the chiral kinetic equation~(\ref{eq:offshell_CKE}) under the on-shell condition $\tilde{q}^2=\hbar S^{\mu\nu}_{\tilde{q}}(\tilde{\Delta}_{\mu}\tilde{q}_{\nu})$.

\section{Feynman propagators for chiral fermions}\label{app:FeynmannW}
The lesser and greater Wigner functions for chiral fermions are given by \cite{Hidaka:2016yjf,Hidaka:2017auj} 
\begin{eqnarray}\label{WF_L_full}
	\mathcal{W}^{\lessgtr \mu}_{\chi}(q)=2\pi{\rm sgn }(q\cdot n)\left[\delta(q^2)\big(q^{\mu}+ \chi\hbar S^{\mu\nu}_{q}\mathcal{D}_{\nu}\big) + \chi\hbar \tilde{F}^{\mu\nu}q_{\nu}\delta'(q^2)\right]f_{\chi}^{\lessgtr}\,,
\end{eqnarray} 
where $f_{\chi}^<=f_{\chi}$ and $f_{\chi}^>=1-f_{\chi}$, $\mathcal{D}_{\mu}f_{\chi}=\Delta_{\mu}f_{\chi}-\Sigma^<_{\chi\mu}f^>_{\chi}+\Sigma^>_{\chi\mu}f^<_{\chi}$, and $S^{\mu\nu}_q=\epsilon^{\mu\nu\alpha\beta}q_{\alpha}n_{\beta}/(2q\cdot n)$. The Feynman propagator $\mathcal{W}^{\mu}_{\chi}(q)$ is given by
\begin{eqnarray}
	\mathcal{W}^{\mu}_{\chi}(q)=\int^{\infty}_{-\infty}\rd q'_0 \left[\tilde{\theta}(q'_0-q_0)\mathcal{W}^{> \mu}_{\chi}(q')-\tilde{\theta}(q_0-q'_0)\mathcal{W}^{< \mu}_{\chi}(q')\right]_{q'^i=q^i}\,,
\end{eqnarray}
where $q_0=q\cdot n$ and 
\begin{eqnarray}
	\tilde{\theta}(q)=\frac{1}{2\pi}\left[\frac{1}{\ri q}+\pi\delta(q)\right]
\end{eqnarray}
corresponds to the Fourier transform of the unit step function. One can accordingly rearrange $\mathcal{W}^{\mu}_{\chi}(q)$ into the form
\begin{eqnarray}
	\mathcal{W}^{\mu}_{\chi}(q)=-\ri\int^{\infty}_{-\infty}\frac{\rd q'_0}{2\pi}	\frac{1}{q'_0-q_0}\left[\mathcal{W}^{> \mu}_{\chi}(q')+\mathcal{W}^{< \mu}_{\chi}(q')\right]_{q'^i=q^i}
	+\frac{1}{2}\left[\mathcal{W}^{> \mu}_{\chi}(q)-\mathcal{W}^{< \mu}_{\chi}(q)\right]\,.
	\nonumber \\
\end{eqnarray}

We may now make the decomposition $\mathcal{W}^{\mu}_{\chi}(q)=\mathcal{W}^{\rm (0)\mu}_{\chi}(q)+\hbar \mathcal{W}^{\rm (1) \mu}_{\chi}(q)$ and $f_{\chi}=f^{(0)}_{\chi}+\hbar f^{(1)}_{\chi}$, up to $\mathcal{O}(\hbar)$. For the classical contribution $\mathcal{W}^{\rm (0)\mu}_{\chi}(q)$, we find \cite{Bellac:2011kqa}
\begin{align}\nonumber
	\mathcal{W}^{\rm (0)\mu}_{\chi}(q)&=-\ri\int^{\infty}_{-\infty}\rd q'_0	\frac{ q'^{\mu}\delta(q'^2_0-|\bm q|^2)}{q'_0-q_0}\text{sgn}(q'_0)
	+\pi q^{\mu}\delta(q^2)\text{sgn}(q_0)\left[1-2f^{(0)}_{\chi}(q)\right]
	\\\nonumber
	&=\frac{\ri q^{\mu}}{q^2}
	+\pi q^{\mu}\delta(q^2)\text{sgn}(q_0)\left[1-2f^{(0)}_{\chi}(q)\right]
	\\
	&=\frac{\ri q^{\mu}}{q^2+\ri\epsilon}-2\pi q^{\mu}\delta(q^2)\left[\Theta(-q_0)+\text{sgn}(q_0)f^{(0)}_{\chi}(q)\right]\,,
\end{align}
where we used $\text{sgn}(q_0)=1-2\Theta(-q_0)$ and
\begin{eqnarray}
	\frac{1}{q^2+\ri\epsilon}=-\ri\pi\delta(q^2)+\frac{1}{q^2}\,.
\end{eqnarray}

To compute the quantum contribution $\mathcal{W}^{\rm (1) \mu}_{\chi}(q)$, we further decompose $\mathcal{W}^{\rm (1) \mu}_{\chi}(q)=\mathcal{W}^{\rm (1a) \mu}_{\chi}(q)+\mathcal{W}^{\rm (1b) \mu}_{\chi}(q)$, where
\begin{align}
	\mathcal{W}^{\rm (1a) \mu}_{\chi}(q)&\equiv -\ri\int^{\infty}_{-\infty}\frac{\rd q'_0}{2\pi} \frac{1}{q'_0-q_0}\left[\mathcal{W}^{(1)> \mu}_{\chi}(q')+\mathcal{W}^{(1)< \mu}_{\chi}(q')\right]_{q'^i=q^i}\,,
 \\
	\mathcal{W}^{\rm (1b) \mu}_{\chi}(q)&\equiv \frac{1}{2}\left[\mathcal{W}^{(1)> \mu}_{\chi}(q)-\mathcal{W}^{(1)< \mu}_{\chi}(q)\right]\,.
\end{align}
From Eq.~(\ref{WF_L_full}), one finds
\begin{align}
	\mathcal{W}^{\rm (1a) \mu}_{\chi}(q)&= -\chi\frac{\ri}{2}\int^{\infty}_{-\infty}\rd q'_0 \frac{{\rm sgn }(q'_0)}{q'_0-q_0} \tilde{F}^{\mu\nu}\partial_{q'{\nu}}\delta(q'^2)\Big|_{q'^i=q^i}\,, \\
	\mathcal{W}^{\rm (1b) \mu}_{\chi}(q)&=-\pi{\rm sgn }(q_0)\left(2\delta(q^2)\left[q^{\mu}f^{(1)}_{\chi}(q)+\chi  S^{\mu\nu}_{q}\mathcal{D}_{\nu}f^{(0)}_{\chi}(q)\right]
    +\chi \tilde{F}^{\mu\nu}q_{\nu}\delta'(q^2)\left[2f^{(0)}_{\chi}(q)-1 \right]\right)\,.
\end{align}
Performing the integration by parts and dropping the surface terms, $\mathcal{W}^{\rm (1a) \mu}_{\chi}(q)$ becomes
\begin{align}\nonumber\label{eq:W1a_in_B}
	\mathcal{W}^{\rm (1a) \mu}_{\chi}(q)&= \ri\chi B^{\mu}\int^{\infty}_{-\infty}\rd q'_0	\frac{2(q'_0-q_0)\delta(q_0')-{\rm sgn }(q'_0)}{2(q'_0-q_0)^2} \delta(q'^2_0-|\bm q|^2)
	\\
	&=-\ri\chi B^{\mu}\left[\frac{\delta(|\bm q|)}{2|\bm q|(q_0-|\bm q|)}+\frac{q_0}{(q^2_0-|\bm q|^2)^2}\right]\,,
\end{align} 
where $B^{\mu}=\tilde{F}^{\mu\nu}n_{\nu}$. Nonetheless, the term proportional to $\delta(|\bm q|)$ in Eq.~(\ref{eq:W1a_in_B}) is unphysical since $q_0=|\bm q|=0$ is at the infrared region beyond the validity of the $\hbar$ expansion, and we shall accordingly drop it. 

When $E^{\mu}=F^{\mu\nu}n_{\nu}=0$ and $f_{\chi}$ only depends on $q_0$, we find 
\begin{eqnarray}
\label{eq:W_general}
	\mathcal{W}^{\mu}_{\chi}(q)=\frac{\ri}{q^2}\left(q^{\mu}-\frac{\chi\hbar q_0B^{\mu}}{q^2}\right)
	+\pi \left[q^{\mu}\delta(q^2)+\chi\hbar B^{[\mu}n^{\nu]}q_{\nu}\delta'(q^2)\right]\text{sgn}(q_0)\left[1-2f_{\chi}(q)\right]
	\nonumber \\
\end{eqnarray}
up to $\mathcal{O}(\hbar)$. In equilibrium at finite temperature $T$ and chemical potential $\mu$, one may introduce 
\begin{align}
\tilde f_{\chi}(q)=\frac{\Theta(q_0)}{\re^{\beta(|q_0|-\mu)}+1}+\frac{\Theta(-q_0)}{\re^{\beta(|q_0|+\mu)}+1}\,,
\end{align}
which satisfies the property $\bar f_{\chi}(q)={\rm sgn}(q_0)\tilde f_{\chi}(q)+\Theta(-q_0)$. Then, Eq.~(\ref{eq:W_general}) can be rewritten as
\begin{align}
	\mathcal{W}^{\mu}_{\chi}(q)=\frac{\ri q^{\mu}}{q^2+\ri\epsilon}
	-\chi\hbar \left[\frac{\ri q_0B^{\mu}}{(q^2)^2}+\pi B^{[\mu}n^{\nu]}q_{\nu}\delta'(q^2)\right]
	-2\pi  \left[q^{\mu}\delta(q^2)+\chi\hbar B^{[\mu}n^{\nu]}q_{\nu}\delta'(q^2)\right]\tilde f_{\chi}(q)\,.
\end{align}

\bibliography{CKT_self-energy_v3.bbl}
\end{document}